\documentclass[useAMS,usenatbib,a4paper]{mn2e}
\usepackage{amssymb,amsmath}
\usepackage{graphicx}
\usepackage{natbib}
\usepackage{journal_shortcuts}

\begin{document}

\title{Simulating the Toothbrush: Evidence for a triple merger of galaxy clusters}
\author[M. Br\"uggen, R.~J. van Weeren, H.~J.~A. R\"ottgering]{M. Br\"uggen$^{1,2}$\thanks{E-mail: mbrueggen@hs.uni-hamburg.de}, R.~J. van Weeren$^{3,4}$, H.~J.~A. R\"ottgering$^{3}$\\ \\
$^{1}$Jacobs University Bremen, P.O. Box 750561, 28725 Bremen, Germany\\
$^{2}$ Hamburger Sternwarte, Universit\"at Hamburg, Gojenbergsweg 112, 21029, Hamburg, Germany\\
$^{3}$Leiden Observatory, Leiden University, P.O. Box 9513, NL-2300 RA Leiden, The Netherlands\\
$^{4}$ Netherl. Foundation for Research in Astronomy (ASTRON), P.O. Box 2, 7990 AA Dwingeloo, The Netherlands}
\maketitle

\label{firstpage}

\begin{abstract}
The newly discovered galaxy cluster 1RXS J0603.3+4214 hosts a 1.9 Mpc long, bright radio relic with a peculiar linear morphology. Using hydrodynamical N-body AMR simulations of the merger between three initially hydrostatic clusters in an idealised setup, we are able to reconstruct the morphology of the radio relic. Based on our simulation, we can constrain the merger geometry, predict lensing mass measurements and X-ray observations. 
Comparing such models to X-ray, redshift and lensing data will validate the geometry of this complex merger which helps to constrain the parameters for shock acceleration of electrons that produces the radio relic.
\end{abstract}


\section{Introduction}

\begin{figure}
\begin{center}
\includegraphics[width=\columnwidth]{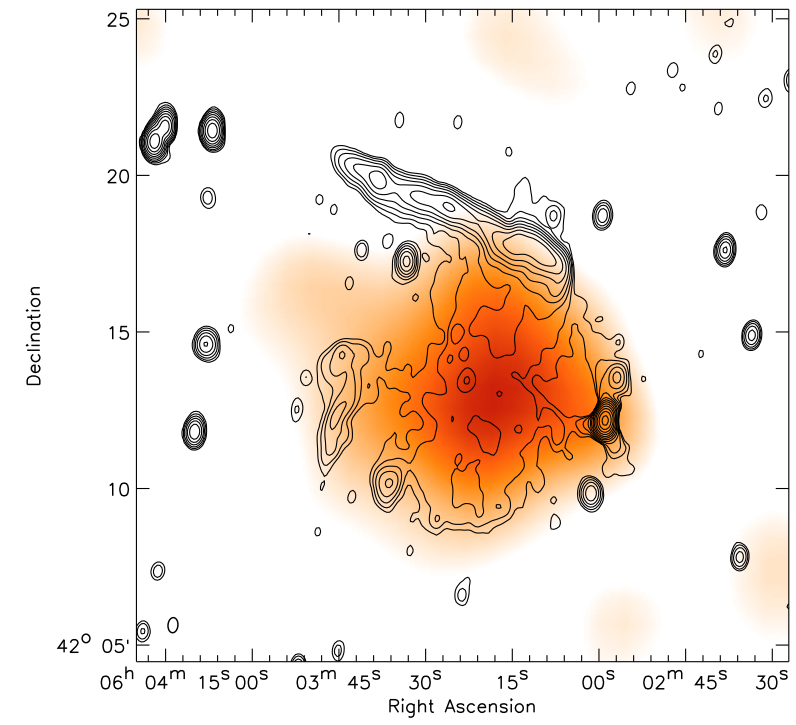}
\caption{Radio - X-ray overlay of 1RXS J0603.3+4214. The image from the ROSAT All Sky Survey was smoothed with a 200'' FWHM Gaussian and is shown 
in orange colors. Solid contours are from the WSRT L-band image and drawn at levels of [1, 2, 4, 8, . . .] $\times$ 0.15 mJy/beam. From van Weeren et al. (2012).}
\label{fig:tooth}
\end{center}
\end{figure}

Some merging galaxy clusters host diffuse extended radio emission, so-called radio halos and relics, unrelated to individual galaxies. 
The origin of these halos and relics is still debated, although there is compelling evidence that they are related to cluster mergers 
\citep{ensslin98, miniati01, hoeft07, hoeft08, pfrommer08, battaglia09, skillman11, vazza12}. Van Weeren (2012) present detailed Westerbork Synthesis Radio Telescope (WSRT) and Giant Metrewave Radio Telescope (GMRT) radio 
observations between 147 MHz and 4.9 GHz of a new radio-selected galaxy cluster 1RXS J0603.3+4214, located at a redshift 
of 0.225. The cluster hosts a large bright 1.9 Mpc radio relic, so-called toothbrush for its shape, as well as an elongated $\sim 2$ Mpc radio halo, and two fainter radio relics (van Weeren et al. 2012 submitted). Part of the 
main radio relic has a peculiar linear morphology. The cluster is detected as an extended X-ray source in the ROSAT All Sky Survey 
with an X-ray luminosity of $L_{X, 0.1-2.4 {\rm keV}}\sim 1\times10^{45}$ erg s$^{-1}$. An overlay of the radio and X-ray data is shown in Fig.~\ref{fig:tooth}. 
The distorted intracluster medium (ICM) is clearly elongated along the N-S and possibly in the E-W directions, indicating a  
complex merger. V,R, and I band 4.2m William Herschel Telescope (WHT) images confirmed the presence of a large galaxy cluster.

For the bright radio relic a clear spectral index gradient from the front of the relic towards the cluster centre is observed. Parts of the relic are highly polarized with a 
polarization fraction of up to 60\%. A model in which particles are (re)accelerated in a first-order Fermi process at the front of the 
relic provides the best match to the observed spectra. The orientation of the bright relic and halo indicate a 
north-south merger, but the peculiar linear shape and the presence of another relic, perpendicular to the bright relic, suggest a 
more complex merger. Deep X-ray observations as well as lensing studies will be needed to completely understand the dynamics of this cluster. Here we attempt to model this peculiar object using the hydrodynamic AMR code FLASH v3.3. In particular, we would like to model (i) the linear morphology of the relic and (ii) its orientation with respect to the X-ray emission. As a result, we will be making predictions about the mass profile that may be detected via gravitational lensing. Understanding the dynamics of the gas in this object is crucial for understanding the physical conditions under which relics are formed.
There exist a number of simulations of cluster mergers \citep{1993ApJ...407L..53R,1993A&A...272..137S, 1994ApJ...427L..87B, 1994MNRAS.268..953P, 1997ApJS..109..307R, 1998ApJ...496..670R,1999ApJ...518..594R,2000ApJ...538...92R, 2001ApJ...561..621R, 2000ApJ...535..586T, 2002MNRAS.329..675R, 2006MNRAS.373..881P,2009ApJ...699.1004Z, 2010ApJ...717..908Z, 2011ApJ...728...54Z, vanweeren11}. 

Cluster mergers can decouple the baryonic matter component from the dark matter (DM) which causes an offset between the gravitational centre (measured from lensing)  and X-ray centre of the cluster. This was first observed for the ``Bullet cluster''  \citep[1E0657$-$56,][]{2006ApJ...648L.109C}. \cite{2007MNRAS.380..911S} presented hydrodynamical models of galaxy cluster mergers to reproduce the dynamical state and mass models (from gravitational lensing) of the ``Bullet'' cluster (1E0657$-$56). \cite{2008MNRAS.389..967M} presented detailed  N-body/SPH simulations of the system. In the following, we describe our attempts to reproduce the event that could form the "toothbrush cluster". 
\vspace{-0.5cm}

\section{Method}

We performed our simulations using FLASH3.3, a parallel 
hydrodynamics simulation code developed at the Center for Astrophysical Thermonuclear Flashes 
at the University of Chicago \citep{fryxell00}. FLASH solves the Euler equations of hydrodynamics on an adaptive, block-structured grid using the Piecewise-Parabolic Method (PPM) of \cite{colella84}, which 
is ideally suited for capturing shocks.  In our simulations, the grid was refined adaptively using standard criteria based on density gradients. FLASH computes the motions of the particles using a particle-mesh method to calculate the gravitational forces. 
We set up three idealised galaxy clusters in a cubical box of size (9 Mpc)$^3$, in our fiducial run resolved with an effective resolution of (512)$^3$ cells. This yields a minimum cell size of 17.5 kpc. The dark matter in each cluster is represented by $10^6$ particles of mass.
For simplicity we neglected cooling and the cosmological expansion of space. The gas is assumed to be ideal with a polytropic equation of state with a ratio of specific heat of 5/3.

\subsection{Initial Conditions}

For simplicity, we follow \cite{zuhone09} and use a Hernquist profile \citep{her90} for the dark matter density distribution:

\begin{equation}
\rho_{\rm DM}(r) = \rho_s\frac{1}{r/a(1+r/a)^3}, 
\end{equation}
where $\rho_s = M_0/(2{\pi}a^3)$ is the scale density of the profile. The Hernquist mass profile converges to $M_0$ as $r \rightarrow \infty$:

\begin{equation}
M(r) = M_0{\frac{r^2}{(r+a)^2}} \, .
\end{equation}

This form of the dark matter density profile is chosen because (i) it resembles the Navarro-Frenk-White (NFW) profile in that as $r \rightarrow 0$, $\rho(r) \propto r^{-1}$, and (ii) the corresponding distribution functions can be written in a simple form. Moreover, the initial velocity distribution of the DM is assumed to be isotropic, i.e. $\sigma_\theta = \sigma_\phi$ and the positions and velocities for the dark matter particles are set up as outlined in \citet{kaz06}. For the particle positions, a random deviate $u$ is uniformly sampled in the range [0,1] and the function $u = M_{\rm DM}(r)/M_{\rm DM}(r_{\rm max})$ is inverted to give the radius of the particle from the centre of the halo. The distribution function of any steady-state, spherically symmetric system depends on the phase space coordinates only via the integrals of motion $E$ and $L$. The relative energy can be written as $E = \psi - \frac{1}{2}v^2$. $L = rv_t$ is the angular momentum of a particle \citep{bin87}, $\psi(r) = -\phi(r)$ is the relative gravitational potential, and $v_t$ is the tangential velocity.  We initialise particle velocities by choosing $L=0$, since cosmological simulations point to low spin parameters for cluster-sized dark matter halos (e.g. \citet{gottloeber07}).

The total mass distribution determines the gravitational potential, and via the constraint of hydrostatic equilibrium of the gas within this potential, the pressure profile is uniquely determined. To obtain densities, we have to impose a gas temperature profile. We assume an isothermal cluster with the temperature set to the virial temperature. The normalisation of the density is given by the cosmic baryon fraction.
We experimented with a number of setups of three-way mergers between clusters, varying the masses and initial positions of the clusters. Here we present a simple setup of three clusters that reproduce some of the morphological features observed in 1RXS J0603.3+4214. See Table 1 for a summary of the cluster parameters.  The relative velocities between the clusters are close to the virial velocities. Thus, this setup is different from that of the bullet cluster which requires much bigger velocities that are much rarer in cosmological simulations \citep{hayashi06}.

A more detailed sampling of the parameter space only makes sense, once further X-ray and lensing studies have yielded more stringent constraints.
\vspace{-0.5cm}

\begin{table}
\caption{Simulation parameters}
\begin{center}
\begin{tabular}{|c|c|c|c|c|c|}
 & $\frac{M}{M_{\odot}}$ &  $\frac{a}{\rm kpc}$ & $\frac{kT}{\rm keV}$ & $(x,y,z)/$Mpc & $(v_x, v_y, v_z)/$km/s \\ \hline\\
1 &$5\times 10^{14}$ & 500 & 3 & (0.65,1.3, 0) & (0,-750, 0)\\
2 &$5\times 10^{14}$ & 500 & 3 & (0.65,-1.3, 0) & (0,750, 0)\\
3& $3.5\times 10^{13}$ & 300 & 3 & (1.95, -1.95, 0) & (-1300,0, 0)
\end{tabular}
\end{center}
\label{default}
\end{table}%

\section{Results}

\begin{figure*}
\begin{center}
\includegraphics[width=0.35\textwidth]{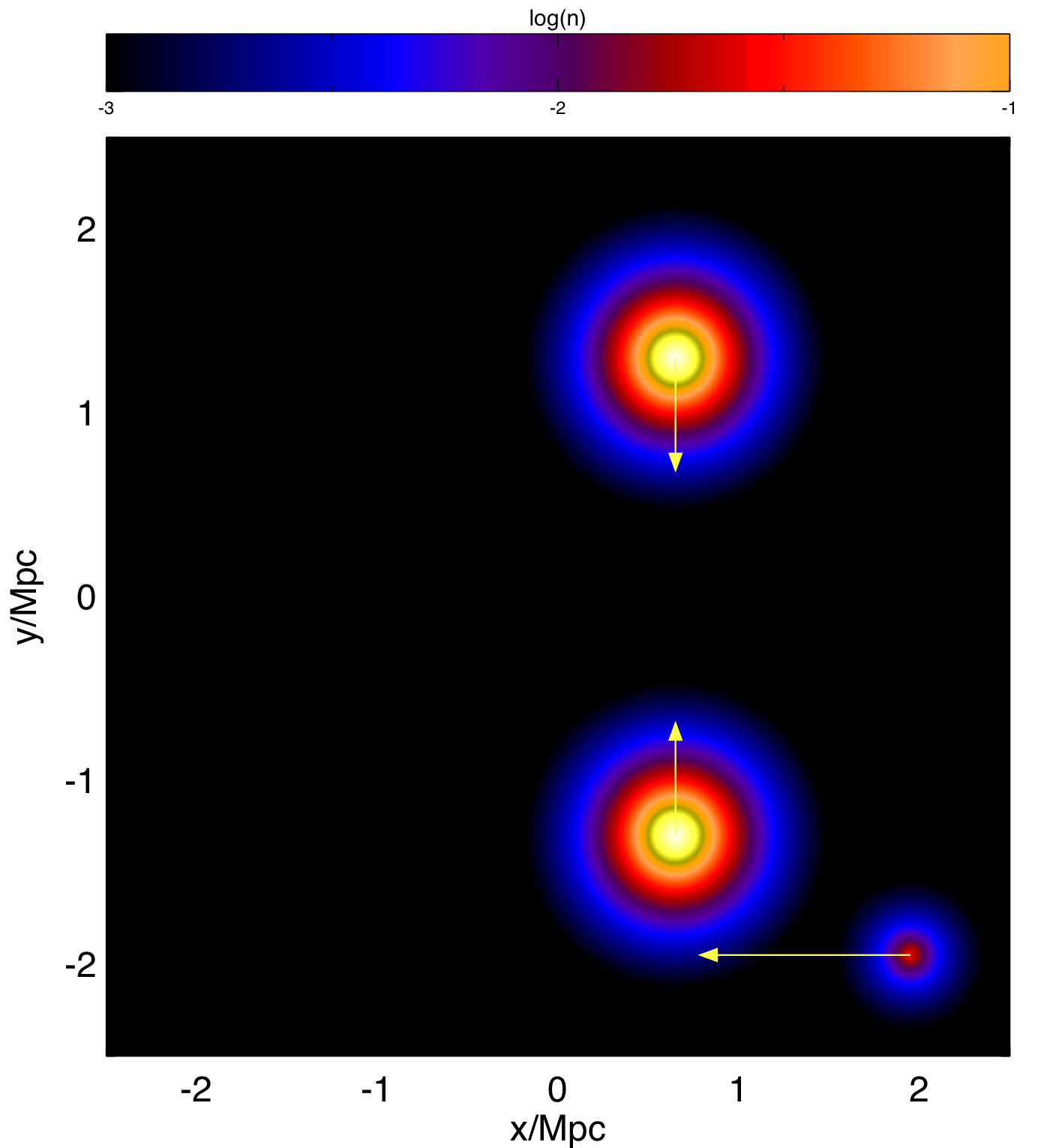}
\includegraphics[width=0.35\textwidth]{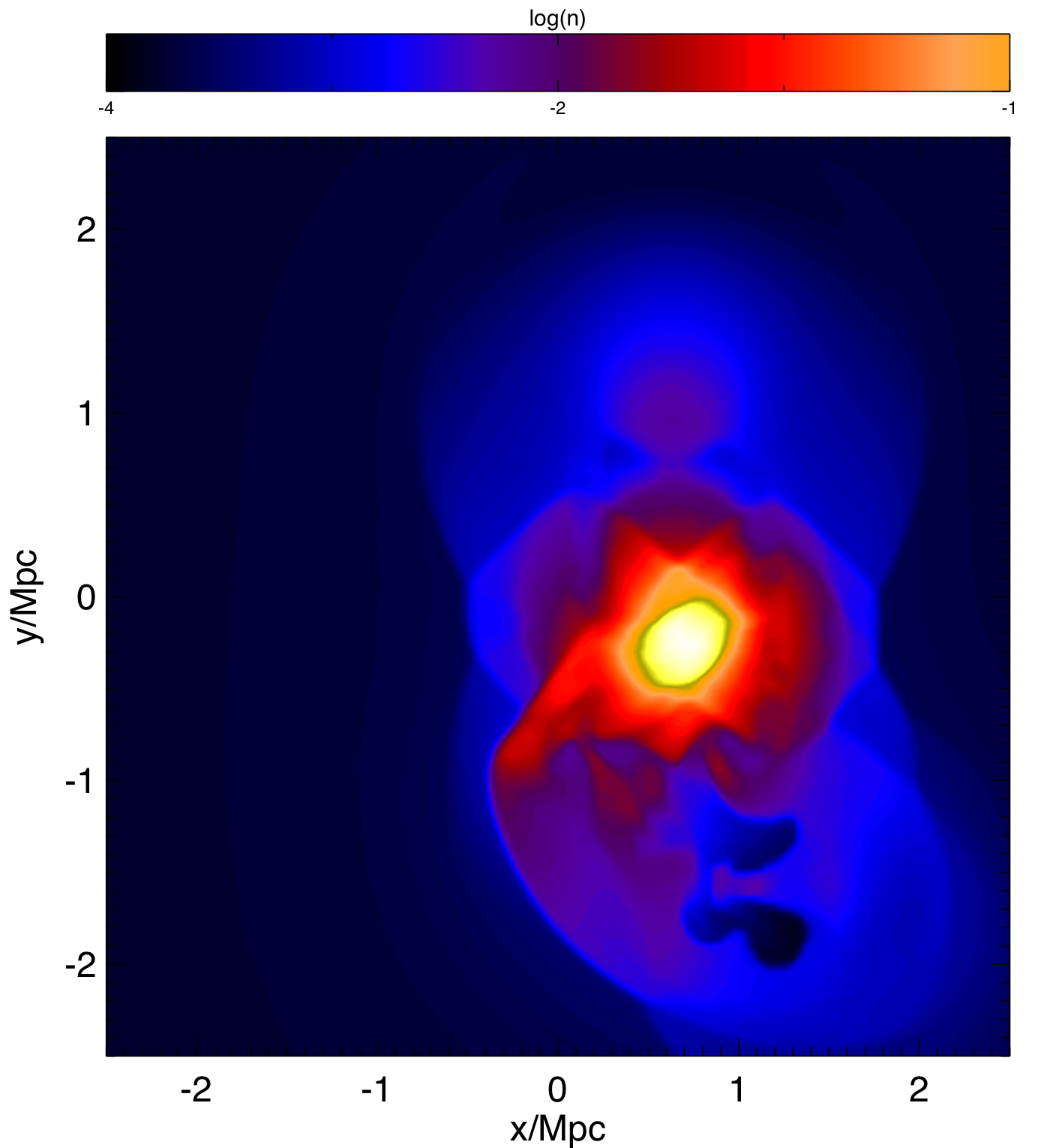}\\
\includegraphics[width=0.35\textwidth]{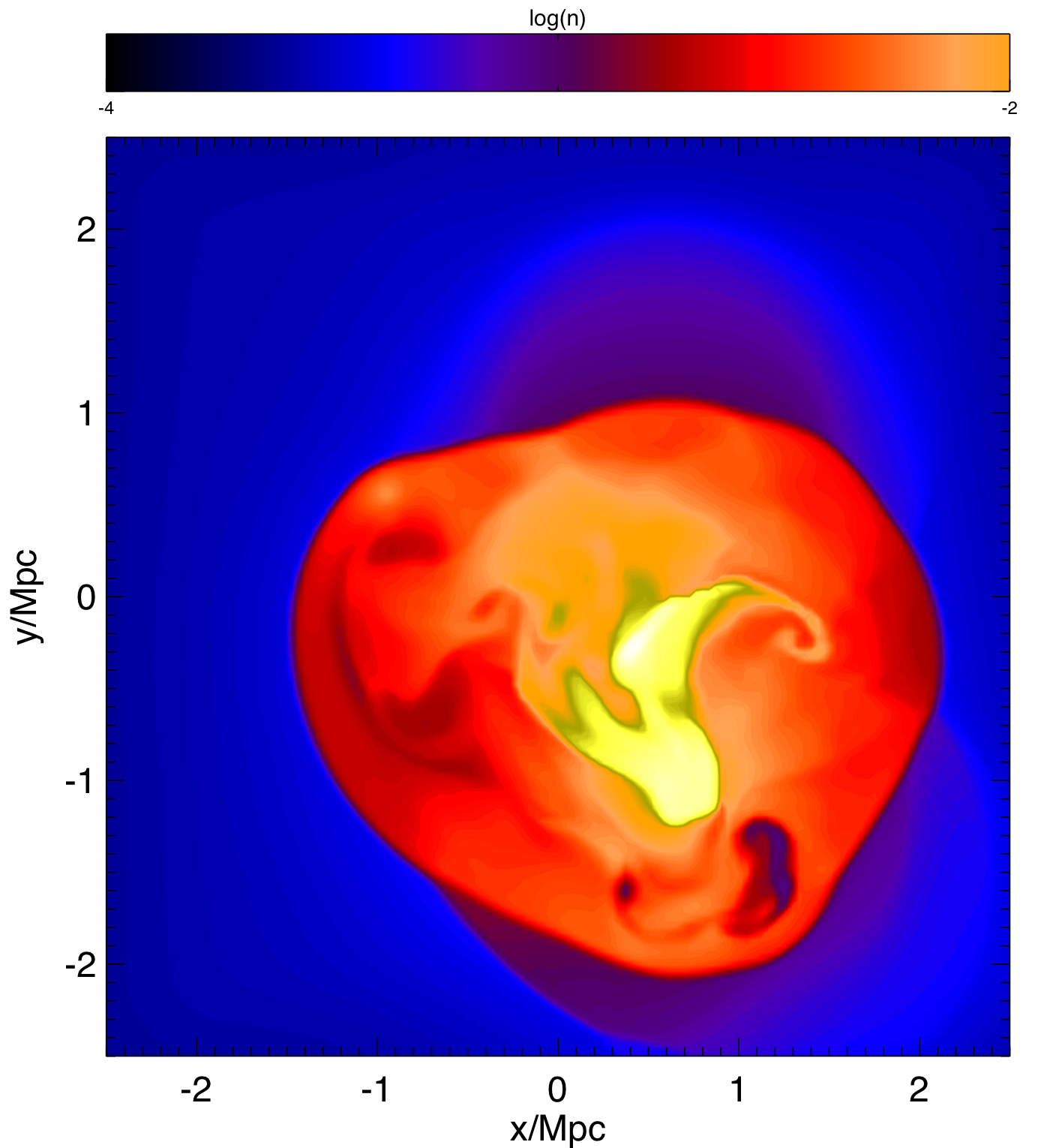}
\includegraphics[width=0.35\textwidth]{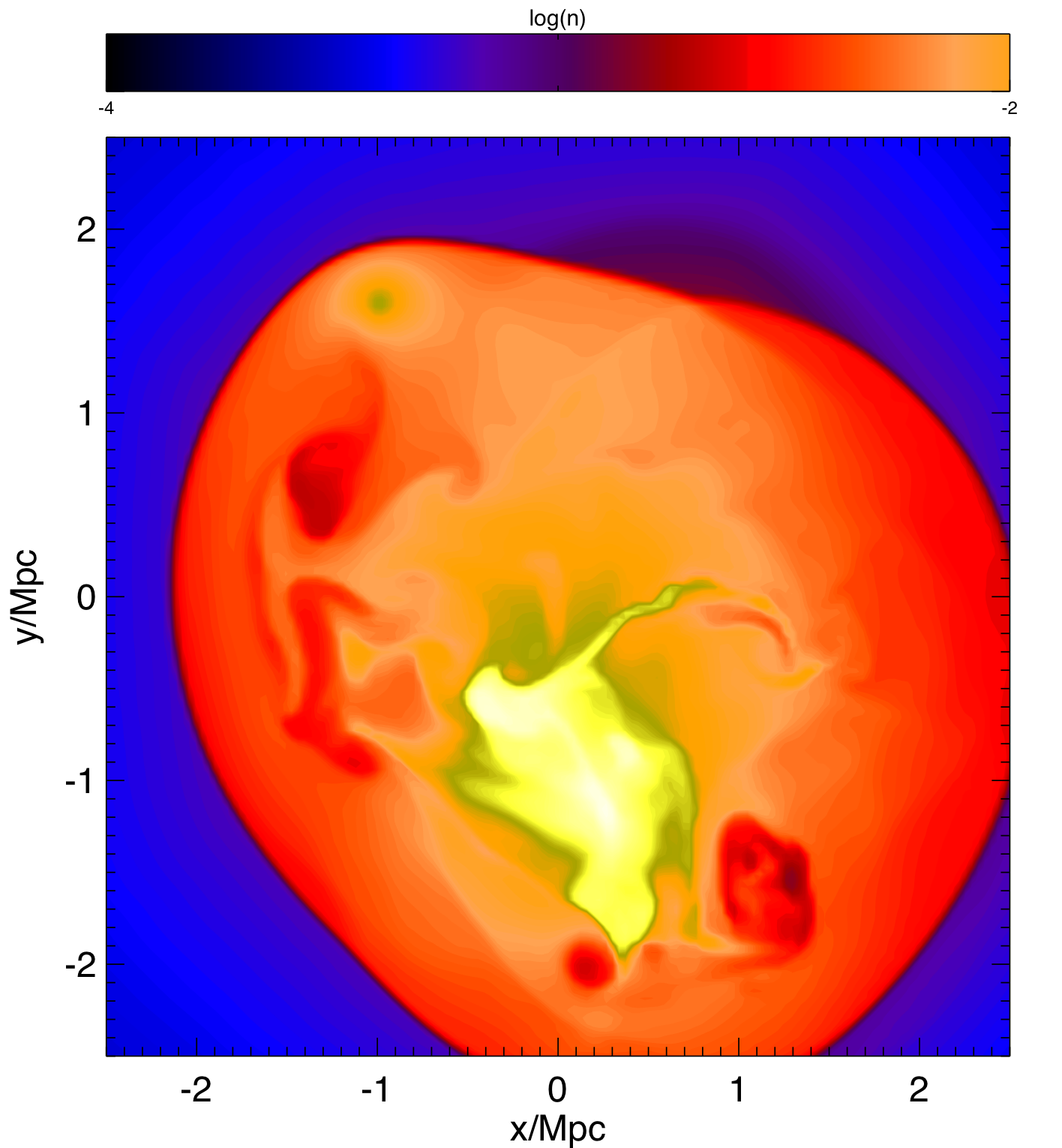}
\caption{Slices of density at times 0, 1.5, 2.5 and 3.5 Gyr after the start of the simulation. Not the entire computational domain is shown.}
\label{fig:dens}
\end{center}
\end{figure*}

First, clusters 1 and 2 that have an initial relative velocity of 1500 km s$^{-1}$ collide with their cores passing at $\sim$ 1.3 Gyr after the start of the simulation.
As clusters 1 and 2 collide in a head-on collision, they drive out an ellipsoidal shock front that is strongest along the merger axis. 
Meanwhile, cluster 3 grazes cluster 2 and loses some of its gas and dark matter. As it then heads north, it is pulled in by the gravity of clusters 1 and proceeds on an arc-shaped orbit. On its course, cluster 3 drives a second major shock into the ICM that merges with the previous shock to form a fairly plane shock front at its near side. This is seen for example in slices of the density that are shown in Fig.~\ref{fig:dens}. In post-processing, we analyse the shocks in the simulation.

\begin{figure*}
\begin{center}

\includegraphics[width=0.35\textwidth]{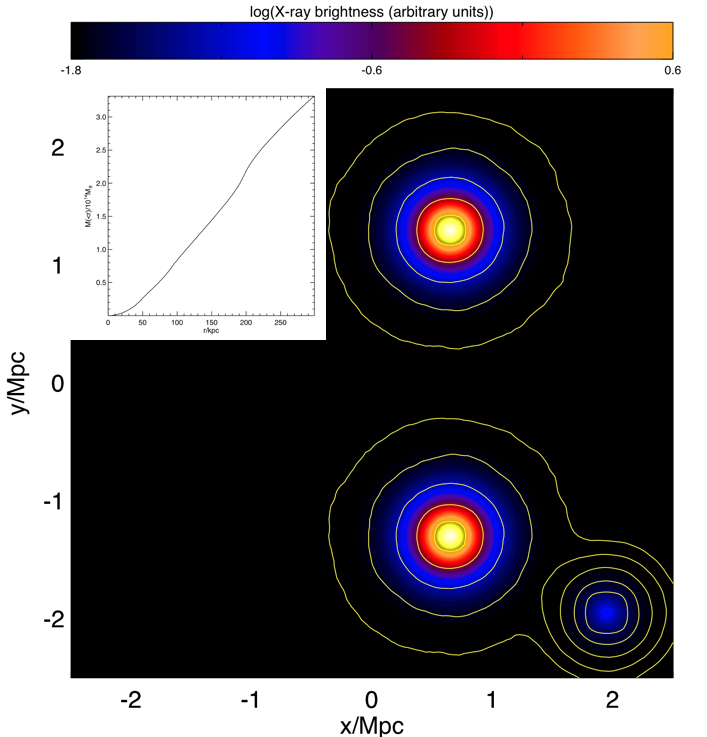}
\includegraphics[width=0.35\textwidth]{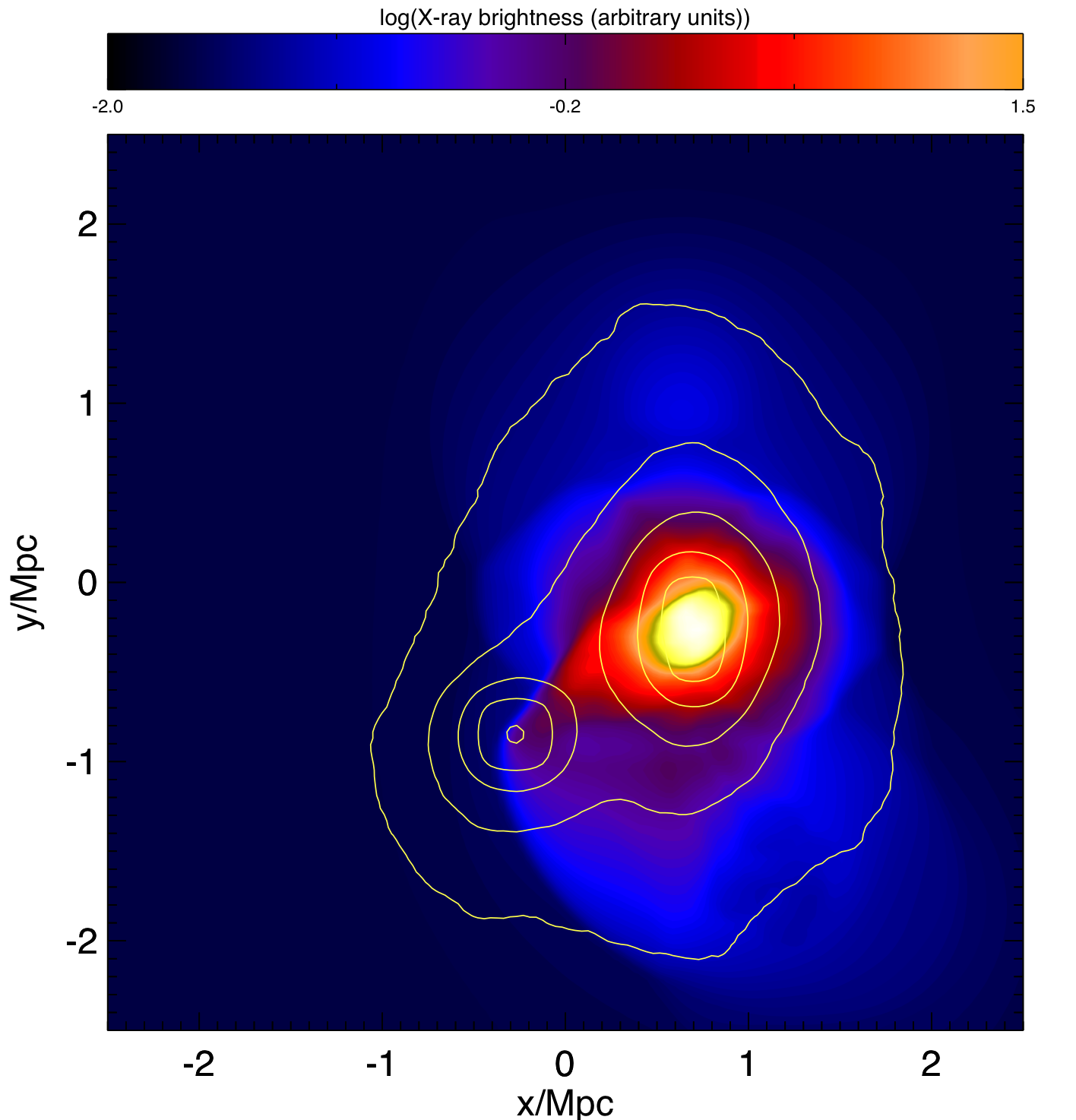}\\
\includegraphics[width=0.35\textwidth]{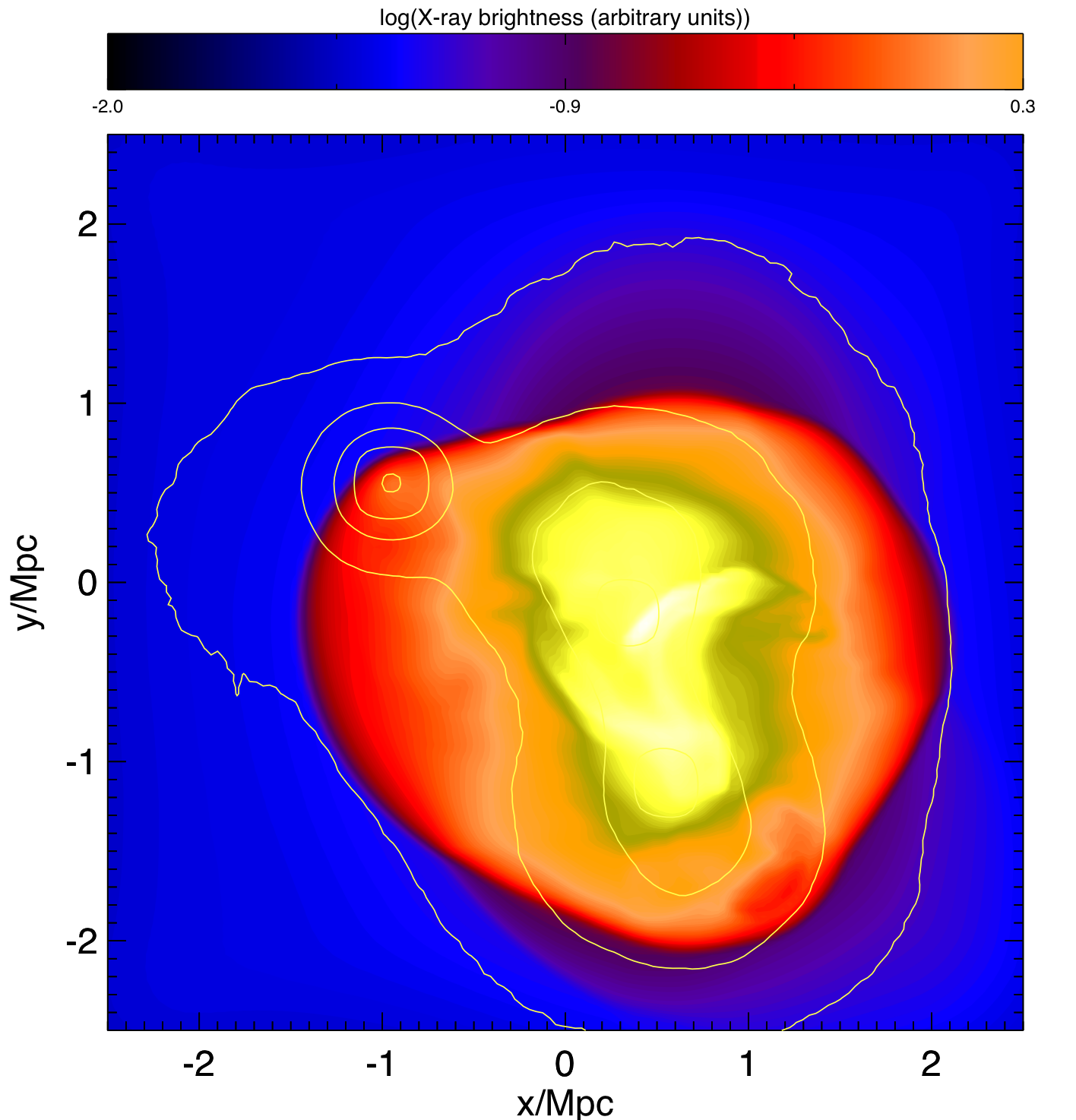}
\includegraphics[width=0.35\textwidth]{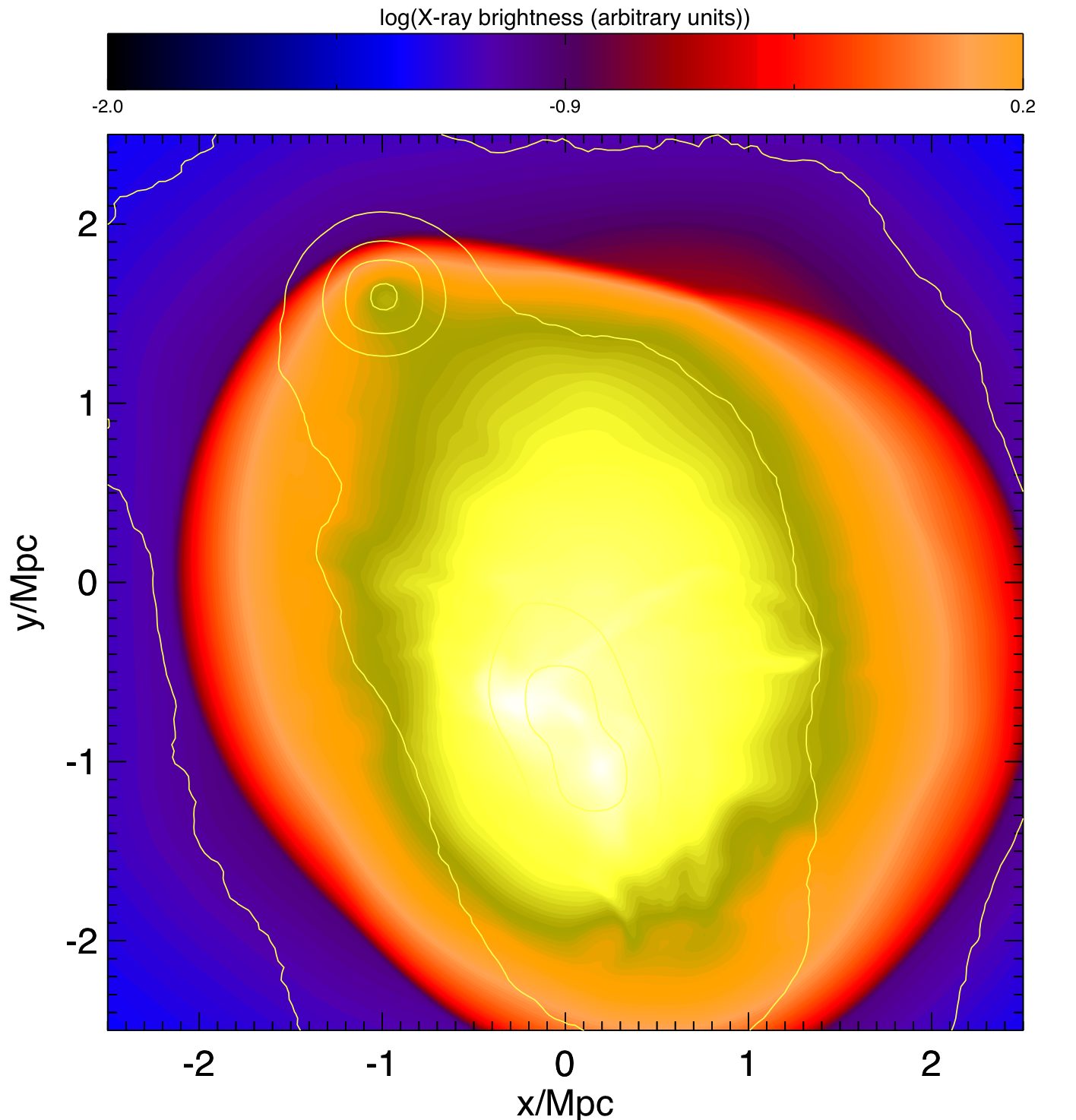}
\caption{Projections of $n^2\sqrt{T}$ as proxy for the X-ray surface brightness at times 0, 1.5, 2.5 and 3.5 Gyr after the start of the simulation. The orange contours trace the surface mass density of the system within $2\times10^3$ and $2\times 10^8 M_{\odot}$ kpc$^{-2}$. Inset in the top left panel is the cumulative projected mass profile at $t=3.5$ Gyr after the start of the simulation. The cylindrical bins are centred on the centre-of-mass of the triple system. }
\label{fig:xray}
\end{center}
\end{figure*}

When cluster 3 collides with the merged clusters 1 and 2, it heats up the temperature of the ICM ahead of it. The hottest temperatures are found in the region around cluster 3. This hot gas drives out a shock that is strongest in the region where the combined shock front forms a region that is straight for $>$ 1 Mpc at the top left of Figs.~\ref{fig:dens} and \ref{fig:xray}. Near the centre-of-mass as well as south of it one sees the cold front from the cold gas from the cluster centres that that is coincident with the dense plume seen in Fig.~\ref{fig:dens}. They have a more complex morphology than those seen in binary mergers.

Projections of $n^2\sqrt{T}$ as proxy for the X-ray surface brightness together with the surface mass density are shown in Fig.~\ref{fig:xray}.
The shocks are visible as discontinuities in the X-ray surface brightness. The bright cores of clusters 1 and 2 separate after core passage with a bright bridge of stripped gas joining them. Also the dark matter halos separate and then finally merge at around 4 Gyr after the start of the simulation. However, the separation between DM and gas (bulleticity) is not nearly as big as in the bullet cluster where the relative velocity between the clusters is much higher. We also computed the projected mass within cylindrical bins centred on the centre-of-mass of the triple system (see inset in Fig.~\ref{fig:xray} which shows the projected mass at 3.5 Gyr after the start of the simulation). We also notice some X-ray bright plumes that extend slighty along the minor axis of the X-ray emission that appear even more pronounced in the ROSAT image. The simulation predicts still a fair amount of X-ray emission at the core of cluster 3, even after 3.5 Gyr. This is not visible in the ROSAT image. If this absence is confirmed this might imply that cluster 3 has less gas than assumed here, possibly as a result of stripping from its first encounter or because it has a higher core entropy than assumed here.

The shocks in our simulation are detected using a multidimensional
shock detection module adopted from the sPPM code (Anderson \&
Woodward 1995) based on pressure jumps across the shock. The basic
algorithm evaluates the jump in pressure in the direction of
compression (determined by looking at the velocity field). If the
total velocity divergence is negative and the relative pressure jump
across the compression front is larger than some chosen value ($\Delta
p/p \ge 0.25$), then a zone is marked as shocked. 

\begin{figure*}
\begin{center}
\includegraphics[width=0.4\textwidth]{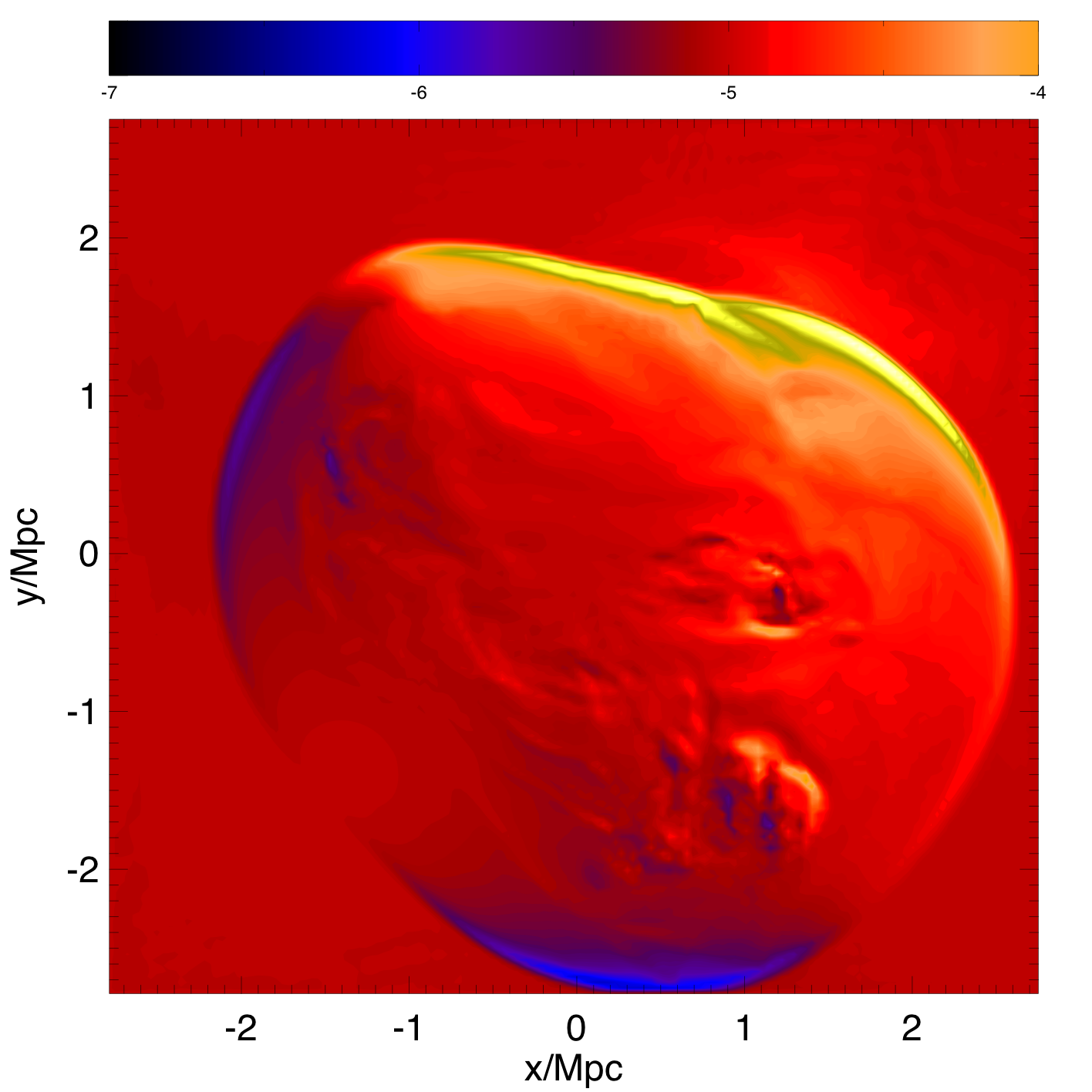}
\caption{Projection of logarithm of the energy dissipated (in units of erg s$^{-1}$cm$^{-2}$) at the shock (at 3.5 Gyrs).}
\label{fig:mach}
\end{center}
\end{figure*}

The projected energy dissipated in the shock through the central regions of our
computational domain are shown in Fig.~\ref{fig:mach}. This may be regarded as a proxy for the energy transferred to cosmic rays, even though the observed radio emission depends on, both, the cosmic ray spectrum as well as the magnetic fields.
The shock strength is highest at the bow of cluster 3 where the Mach number is as high as 4 and the leading edge of the shock forms a fairly straight line. This Mach number is in line with the Mach number derived from the spectral index of the synchrotron emission, which lies between 3.8 and 4.6 (van Weeren et al. 2012 submitted). In accordance with observations, this straight line subtends an angle of roughly 30$^o$ with the normal of the major axis of the X-ray emission. This angle depends on the orbit of cluster three with respect to the main merger. A velocity of cluster 3 that exceeds the propagation speed of the main merger shock causes this oblique angle. Projection of the shock front leads to a width of the dissipated energy contours in Fig.~\ref{fig:mach} of $\sim 200$ kpc. Based on this simulation the radio halo that fills the region between clusters 1 and 2 does not appear to be directly related to a radio relic (produces by a merger shock) that is seen in projection.

We do not model the radio emission as in van Weeren et al (2011) as this would merely introduce more unconstrained parameters. From Fig.~\ref{fig:mach}, we would expect a counter relic. The Mach number is high again in the SE, and one might speculate that the linear structure seen in the radio image in the SE may be indicative of a shock there. However, the orientation does not quite match what we find in the simulation.
While it is impossible to argue for the uniqueness of our reconstruction of this merger, we note that gas clumping as inferred from recent X-ray observations is not able to explain the asymmetry, the angle with respect to the main axis or the straight morphology of the shock front.
In summary, our simulations provide evidence that the relic seen in galaxy cluster 1RXS J0603.3+4214 is the result of a triple merger of clusters. Modelling a merger with mass ratio 1:1:0.07 on a trajectory where the smaller clusters skirts the main head-on collision, we can recover i) the length of the relic, ii), the shape of the relic and iii) the angle with respect to the major axis of the X-ray emission. Upcoming multi-wavelength data will be used to validate this picture.

\vspace{-0.7cm}

\section*{Acknowledgements}

MB acknowledges support by the research group FOR 1254 funded by the Deutsche Forschungsgemeinschaft. RJvW acknowledges funding from the Royal Netherlands Academy of Arts and Sciences. The results presented were produced using the FLASH code, a product of the DOE
ASC/Alliances-funded Center for Astrophysical Thermonuclear Flashes at the University of Chicago.  Finally, we thank the referee for a very constructive report.
\vspace{-0.7cm}

\bibliographystyle{mn2e}
\bibliography{toothbib}

\end{document}